\documentclass{JHEP3}
\usepackage{graphicx}
\usepackage{url}
\def\miss {\hspace{-0.5cm}\slash~~}

\catcode`@=11 
\def \gsim{\mathrel{\mathpalette\@versim>}}
\def \lsim{\mathrel{\mathpalette\@versim<}}
\def \@versim#1#2{\lower0.4ex\vbox{\baselineskip\z@skip\lineskip\z@skip
     \lineskiplimit\z@\ialign{$\m@th#1\hfil##\hfil$%
     \crcr#2\crcr\sim\crcr}}}
\catcode`@=12 

\title{R-parity violating resonant stop production at the Large Hadron
  Collider}

\author{Nishita Desai$^a$ and Biswarup Mukhopadhyaya$^{a,b}$\\
$^a$Regional Centre for Accelerator-based Particle Physics,\\ 
Harish-Chandra Research Institute\\ Chhatnag Road,
 Jhunsi,\\ Allahabad - 211 019, India \\
\\
$^b$Indian Association for Cultivation of Science,\\
Raja SC Mullick Road, \\
Jadavpur, Kolkata - 700032 , India\\
Email: \email{nishita@hri.res.in}, \email{biswarup@hri.res.in}}
\preprint{HRI-RECAPP-2010-002}
\keywords{MSSM, supersymmetry, R-parity violation} 

\abstract{We have investigated the resonant production of a stop at
  the Large Hadron Collider, driven by baryon number violating
  interactions in supersymmetry.  We work in the framework of minimal
  supergravity models with the lightest neutralino being the lightest
  supersymmetric particle which decays within the detector.  We look
  at various dilepton and trilepton final states, with or without
  b-tags. A detailed background simulation is performed, and all
  possible decay modes of the lighter stop are taken into account.  We
  find that higher stop masses are sometimes easier to probe, through
  the decay of the stop into the third or fourth neutralino and their
  subsequent cascades.  We also comment on the detectability of such
  signals during the 7 TeV run, where, as expected, only relatively
  light stops can be probed.  Our conclusion is that the resonant
  process may be probed, at both 10 and 14 TeV, with the R-parity
  violating coupling $\lambda''_{312}$ as low as 0.05, for a stop mass
  of about 1 TeV.  The possibility of distinguishing between resonant
  stop production and pair-production is also discussed.  }

\begin{document}

\section{Introduction}

The current structure of the standard model (SM), with gauge
invariance and renormalisability built in, implies automatic lepton
and baryon number conservation.  This is no longer true in the
supersymmetric (SUSY) extension of the SM
\cite{Nilles:1983ge,Kane:1998hz}, where scalars carrying baryon or
lepton number are present.  Thus the superpotential of the minimal
SUSY standard model (MSSM), namely

\begin{eqnarray}
\label{mssm-supp}
\mathcal{W}_{MSSM} = h^d_{ij} Q_iD^c_jH_d + h^u_{ij} Q_i U^c_j H_u +
h^l_{ij}L_iE^c_jH_d + \mu H_uH_d
\end{eqnarray}

\noindent
can in principle be augmented to include

\begin{eqnarray}
\label{rpv-supp}
\mathcal{W}_{RPV}= \mu_i L_iH_u + \lambda_{ijk}L_iL_jE^c_k +
          \lambda'_{ijk}L_iQ_jD^c_k+ \lambda''_{ijk}U^c_iD^c_jD^c_k
\end{eqnarray}

\noindent which contain terms that are gauge invariant and
renormalisable but explicitly violate lepton or baryon number.  Here,
L(E) is an $SU(2)$ doublet (singlet) lepton superfield and Q (U,D) is
(are) an $SU(2)$ doublet (singlet) quark superfield(s).  $H_u$ and
$H_d$ are the two Higgs doublet superfields, $\mu$ is the Higgsino
parameter and $(i,j,k)$ are flavour indices.  Each term in equation
(\ref{rpv-supp}) violates R-parity, defined as $R=(-1)^{3(B-L)-2S}$
(where B is baryon number, L is lepton number and S is spin), against
which all SM particles are even whereas all superpartners are odd. The
consequence of violating R-parity is that superpartners need not be
produced in pairs anymore, and that the lightest superparticle (LSP)
can now decay.  The strongest argument for studying R-parity violation
is that it does not arise as an essential symmetry of MSSM.  However,
the requirement of suppressing proton decay prompts one to allow {\em
  only one} of B and L to be violated at a time.

The collider phenomenology in the absence of R-parity may be very
different from that of the usual R-parity conserving MSSM. In
particular, if the R-parity violating (RPV) couplings are large
enough, the LSP will decay within the detector and one no longer has
missing-$E_T$ as a convenient discriminator. Although studies have
taken place on such signals, closer looks at them are often quite
relevant in the wake of the Large Hadron Collider (LHC).  In
particular, it is crucial to know the consequences of broken R-parity
in the production of sparticles.  Here we perform a detailed
simulation in the context of the LHC, highlighting one possible
consequence of the B-violating term(s), namely, the resonant
production of a squark---in this case, the stop.

Many of the RPV couplings have been indirectly constrained from
various decay processes, including rare and flavour-violating decays
and violation of weak universality.  The constraints derived are of
two general kinds---those on individual RPV couplings, assuming the
existence of a single RPV term; and those on the products of couplings
when at least two terms are present, which contribute to some (usually
rare) process.  The constraints obtained so far are well-listed in the
literature\cite{Barbier:2004ez}.

The L-violating terms are relatively well-studied, partly because of
their potential role in generating neutrino masses and are constrained
by indirect limits.  In comparison, the baryon-number violating
coupling are relatively unconstrained.  $\lambda''_{112,113}$ are
constrained from double nucleon decay and neutron-antineutron
oscillations\cite{Zwirner:1984is,Dreiner:1991pe,Barbieri:1985ty}.  The
rest of the couplings are constrained only by the requirement that
they remain perturbative till the GUT scale.  Limits on
$\lambda''_{3ij}$ type of couplings due to the ratio of Z-boson decay
widths for hadronic versus leptonic final states have been calculated
for a stop mass of 100 GeV \cite{Bhattacharyya:1995bw}.  However, the
results do not restrict the couplings for high stop masses of concern
here.  The coupling $\lambda''_{3jk}$ is thus practically
unconstrained for large stop masses.  It is also known that mixing in
the quark sector causes generation of couplings of different flavour
structures and can therefore be constrained by data from flavour
changing neutral currents(FCNC)\cite{Agashe:1995qm}.  Such effects
arising from mixing in the quark and squark sector can affect the
contribution of R-parity violation to physical process and alter the
limits\cite{Allanach:1999ic}.  However these effects are model
dependent and have not been taken into account here.

It has been already noticed that such large values of $\lambda''$-type
couplings as are still allowed, not only cause the LSP to decay, but
also lead to resonant production of squarks via quark fusion at the
LHC.  The rate of such fusion can in fact far exceed that of the
canonically studied squark-pair production.  One would therefore like
to know how detectable the resonant process is at the LHC.
Furthermore, one needs to know the search limits in different phases
of the LHC, and how best to handle the backgrounds, both from the SM
and the R-conserving SUSY processes.  These are some of the questions
addressed in this paper.

Single stop production, mostly in the context of the Tevatron, was
studied in detail in\cite{Berger:1999zt,Berger:2000zk}.  A full
one-loop production cross section can be found in\cite{Plehn:2000be}.
A study of SUSY with the LSP decaying through baryon-number violating
couplings and therefore giving no missing energy was done in
\cite{Baer:1994zw}.  Further studies on determining the flavour
structure of baryon number violating couplings and possible mass
reconstruction following specific decay chains can be found
in\cite{Allanach:2001xz,Allanach:2001if}.  There have also been recent
studies on possible LSPs\cite{Dreiner:2008ca} and identification of
R-parity violating decays of the LSP using jet substructure
methods\cite{Butterworth:2009qa}.  A recent study on identification of
stop-pair production via top-tagging using jet-substructure can be
found in \cite{Plehn:2010st}.

We find, however, that the earlier studies on resonant stop production
are inadequate in the context of the LHC.  We improve upon them in the
following respects:
\begin{itemize}
\item In the work done for the Tevatron, the sparticle masses were
  required to be less than 500 GeV to be within reach.  Thus, the
  gluino was also required to be much lighter than a TeV to avoid
  large radiative corrections to squark masses.  This implied that in
  the constrained MSSM (cMSSM) \cite{Kane:1993td} scenario, the LSP,
  assumed to be the lightest neutralino ($\tilde \chi_1^0$), had to
  have mass less than 100~GeV.  If we allow only the term proportional
  to $\lambda''_{3ij}$, the only three-body decay of $\tilde \chi_1^0$
  is $\tilde \chi_1^0 \rightarrow \bar t \bar d_i \bar d_j (tds)$.
  Since the neutralino is much lighter than the top, it can decay only
  via a 4-body decay and therefore has long lifetime and decays
  outside the detector for all allowed values of
  $\lambda''_{3ij}$\cite{Allanach:1999bf}.  Thus, one still has the
  canonical missing-$E_T$ signature.  This was one of the main
  assumptions in\cite{Berger:2000zk}.  However, if the stop mass is
  beyond the Tevatron reach but within the reach of the LHC, we may
  indeed have lightest-neutralino mass high enough to allow decay
  within the detector.
\item We focus on this richer and more challenging scenario within the
  framework of minimal supergravity (mSUGRA) models.  We investigate
  the LHC reach for detection of the lightest stop assuming that the
  lightest neutralino is the LSP which decays within the detector and
  the stop is heavy enough to be beyond the reach of Tevatron.
\item For a light stop, the only available R-parity conserving decay
  modes are into the lightest neutralino($\tilde \chi_1^0$) and
  lighter chargino($\tilde \chi_1^+$) i.e. $\tilde t \rightarrow
  t\tilde \chi_1^0, b\tilde \chi_1^+$.  For stop mass near a TeV, the
  decay modes into higher neutralinos and the heavier chargino may
  open up, leading to different final states.  We have found that this
  drastically improves the detectability of the signature over the SM
  backgrounds.
\item We have taken into account all the potential backgrounds at the
  LHC, including those from $t{\bar t}+jets,Wt{\bar t}+jets,Zt{\bar
    t}+jets$, which pose little problem at the Tevatron. A detailed
  investigation towards reducing these backgrounds has been reported
  in the present study.
\item In identifying signals of stop decay, it is often helpful to tag
  b quarks.  This is especially important since the reconstruction of
  energetic top quarks in multi-top final states is difficult and has
  a low efficiency.  However, the b-jets produced from stop decay can
  be quite hard, especially when the stop is heavy and one is looking
  at the decay in the $b\tilde\chi_1^{+}$ channel. It is not clear
  that the b-tagging efficiency is appreciable at the LHC for b-jets
  with $p_T \gsim$ 100 GeV. With this is mind, we have performed a
  conservative analysis with a b-jet has zero detection efficiency
  unless its $p_T$ lies in the range 50 - 100 GeV\cite{Aad:2009wy}.
\end{itemize}

We present our results for centre of mass energies of 7, 10 and 14 TeV
at the LHC.  In section 2, we discuss the rates for resonant stop
production, the different decays of the stop and our choice of
benchmark points to account for all of them.  In section 3, we present
a detailed description of cuts required to isolate the signal and
section 4 contains the numerical results from our simulation.  We also
comment on the distinguishability of such a signal from dilepton
signals coming from R-parity conserving MSSM. It is possible to have
LSPs other than $\tilde \chi_1^0$ when R-parity is violated.  We
comment on the possibility of detection in such cases in section 5 and
summarise and conclude in section 6.

\section{Resonant stop production and decays}

\subsection {Stop production}
The resonant stop production process depends on B-violating couplings
proportional to $\lambda''_{3ij}$, and also on fraction of the
right-chiral eigenstate (${\tilde t}_R$) in the mass eigenstate
concerned. We concentrate on the production of $\tilde t_1$ since the
lighter stop eigenstate usually has a higher fraction of ${\tilde
  t}_R$.  The resonant production cross section is given by
\begin{eqnarray}
\nonumber \sigma_{\tilde t_1}&=&\frac{2 \pi \sin^2\theta_{\tilde t}}{3 m^2_{\tilde t}} \times \\
& &\sum_{i,j}{ |\lambda''_{3ij}|^2  \int dx_1 dx_2 [f_i(x_1)  f_j(x_2)+f_i(x_2)  f_j(x_1)]  \delta(1-\frac{m_{\tilde t_1}}{\sqrt{\hat s}})}
\end{eqnarray}
where $ \sin\theta_{\tilde t}$ is the amplitude of finding a $\tilde
t_R$ in $\tilde t_1$, $f_i$ is the proton parton distribution function
for a parton of species $i$ and $x_{(1,2)}$ are the momentum fractions
carried by the respective partons.  Out of the three possible
$\lambda''$ couplings, contributions via $\lambda''_{313}$ and
$\lambda''_{323}$ are suppressed due to the small fraction of b quarks
in the proton. We therefore look at the production of top (anti)
squark through the fusion of the d and s (anti)quarks, via the
coupling $\lambda''_{312}$.  Since the actual cross section for
production of the lightest stop depends on the mixing angle via
$\sin^2 \theta_{\tilde t}$, it is useful to define the cross section
in terms of an effective coupling $\lambda''_{eff}=\sin\theta_{\tilde
  t} \lambda''_{312}$.
\begin{eqnarray}
\sigma_{\tilde t_1}& = & \frac{2 \pi}{3 m^2_{\tilde t_1}}  |\lambda''_{eff}|^2  \times  \nonumber\\
& & 2\int dx_1 dx_2  [f_d(x_1)  f_s(x_2) + f_d(x_2)  f_s(x_1)  \nonumber \\
& & +f_{\bar d}(x_1)  f_{\bar s}(x_2)+ f_{\bar d}(x_2)  f_{\bar s}(x_1)] \delta(1-\frac{m_{\tilde t_1}}{\sqrt{\hat s}}) \nonumber \\
\end{eqnarray}

The production cross section at the LHC with centre-of-mass energies
of 7, 10 and 14 TeV is given in Figure~\ref{xsec}.  As an
illustration, we have chosen the value $\lambda''_{eff} =0.2$ which is
consistent with the existing limit on $\lambda''_{312}$.  In general,
both $\tilde t_1$ and $\tilde t_2$ will be produced.  However, due to
larger mass and smaller fraction of $\tilde t_R$ , ${\tilde t_2}$ is
rarely produced.  For comparison, we also present the ${\tilde
  t_1}$-pair production cross-section via strong interaction.  For
$m_{\tilde t_1} > 500$ GeV, the resonant production dominates over
pair-production for $\lambda''_{eff}>0.01$ at 14 TeV.  Resonant
production can therefore hold the key to heavy stop signals if baryon
number is violated.

\FIGURE{
\includegraphics[width=100mm]{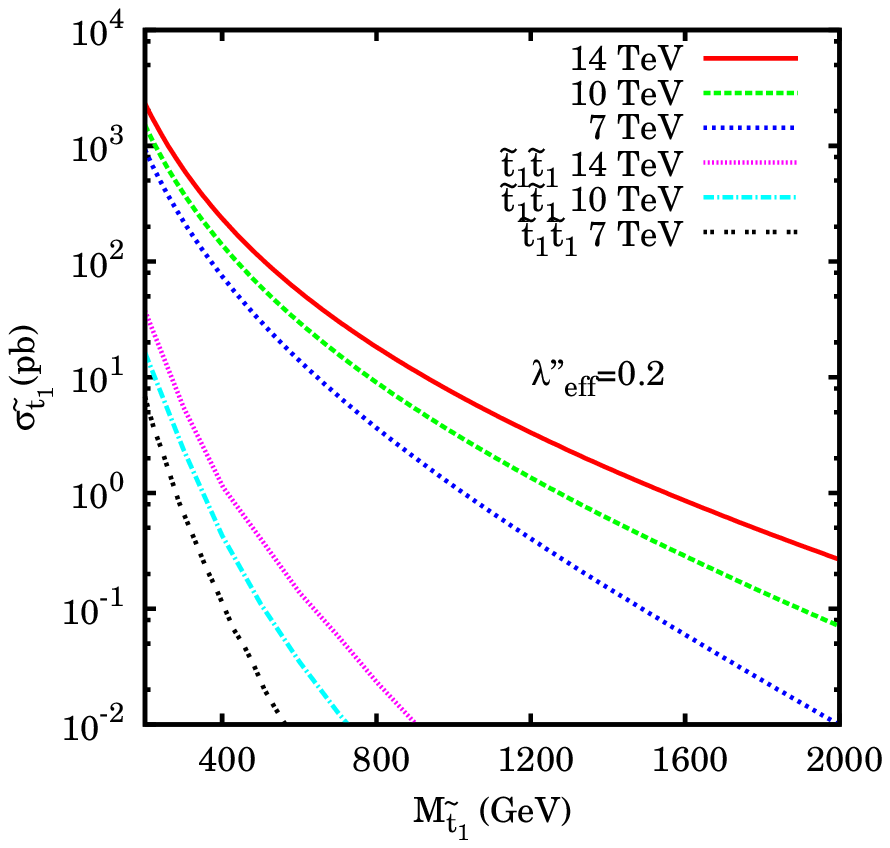}
\caption{Production cross section at the LHC for $\sqrt{s}=7,10$ and
  $14$~TeV with $\lambda''_{eff}=0.2$.  The corresponding cross
  sections for R-conserving pair production are also
  shown.} \label{xsec}
}

At next-to-leading order, the production cross section at
$\sqrt{s}=14$~TeV is modified by a k-factor of about
1.4\cite{Plehn:2000be}.  The uncertainty due to renormalisation and
factorisation scales at lowest order is about 10\% and drops to 5\%
when NLO corrections are taken into account.

\subsection {Stop decays and choice of benchmark points}

\FIGURE{
\includegraphics[width=70mm]{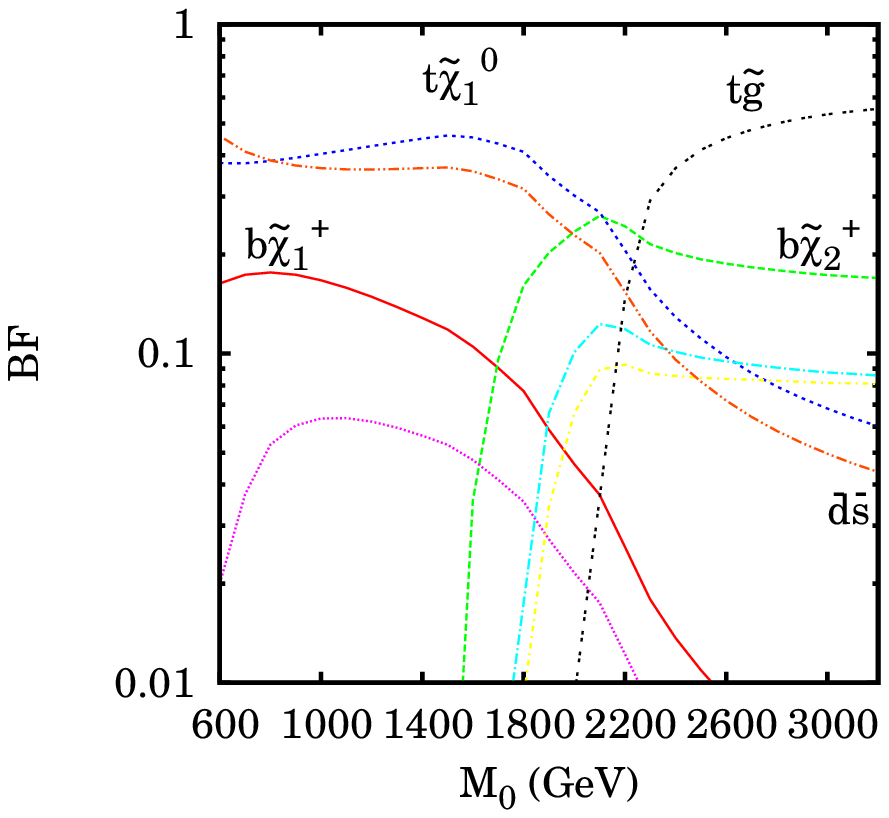}
\includegraphics[width=70mm]{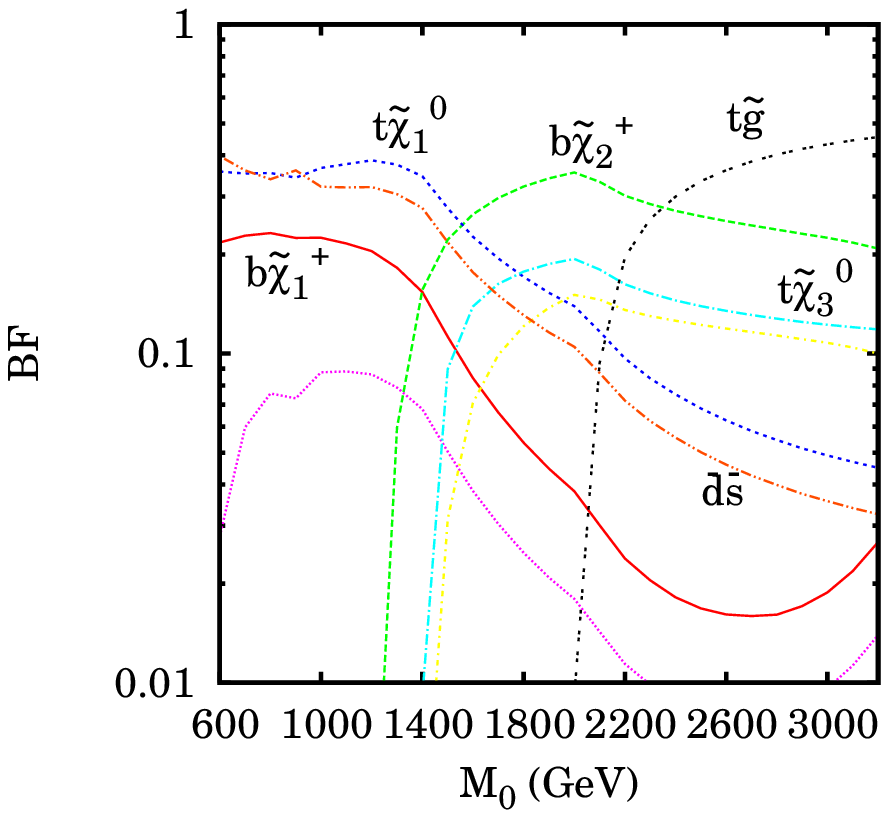}
\includegraphics[width=70mm]{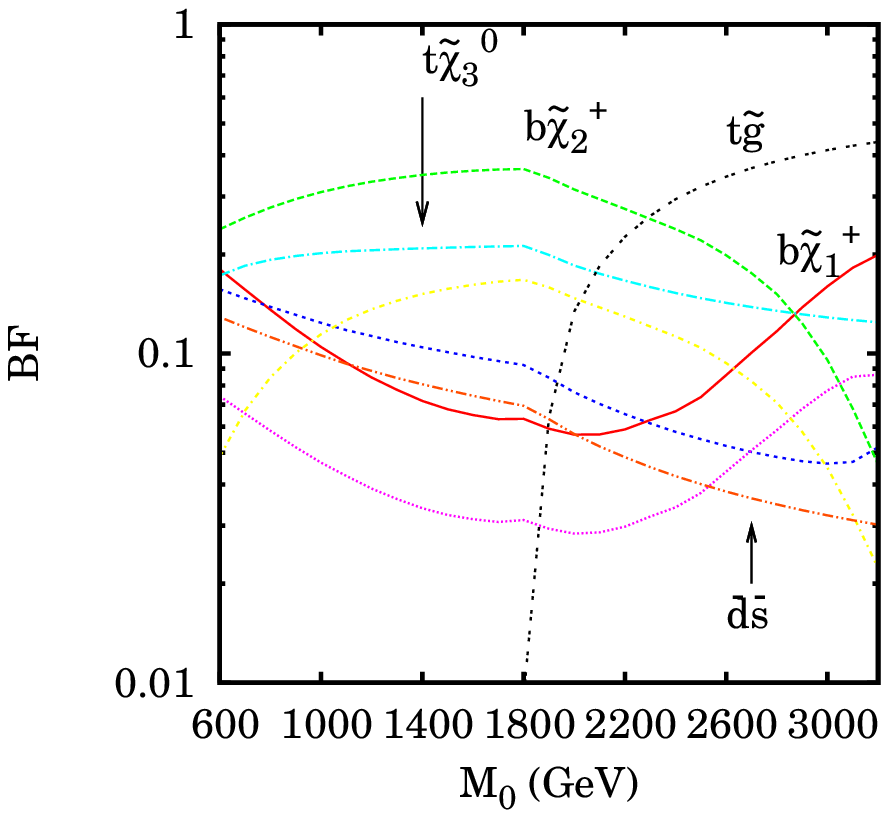}
\caption{Lighter stop decay branching fractions in different modes for
  $\tan \beta=5$, $A_0=-1500$ (top left) ; $\tan \beta=40$,
  $A_0=-1500$ (top right) and $\tan \beta=10$, $A_0=0$
  (bottom). $\lambda''_{312}=0.2$, $\mu>0$ and $m_{1/2}=450$ GeV in
  all cases.}
\label{bf}
}

We wish to make our conclusions apply broadly to a general SUSY
scenario and to include all possible final states arising from stop
decay.  However, the multitude of free parameters in the MSSM often
encourages one to look for some organising principle.  A common
practice in this regard is to embed SUSY in high-scale breaking
scheme. Following this practice, we have based our calculation on the
minimal supergravity (mSUGRA) model\cite{Nath:1983fp}, mainly for
illustrating our claims in a less cumbersome manner. The high scale
parameters in this model are: $m_0$, the unified scalar parameter,
$m_{1/2}$, the unified gaugino parameter, $sign(\mu)$, where $\mu$ is
the Higgsino mass parameter, $A_0$, the unified trilinear coupling and
$\tan \beta$, the ratio of the two Higgs vacuum expectation values.

Although the production cross section of the stop depends only on the
mass and mixing angle of the stop, any strategy developed for seeing
the ensuing signals has to take note of the decay channels. We have
tried to make our analysis comprehensive by including all possible
decay chains of the stop.  Thus we have included decays into $t \tilde
\chi_i^0$, $b\tilde \chi_i^\pm$, $t\tilde g$ and $ds$, of whom the
first three are R-conserving decays while the last one is R-violating.
The charginos, neutralinos or the gluino produced out of stop-decay
have their usual cascades until the LSP (here $\chi_1^0$, the lightest
neutralino) is reached. The $\chi_1^0$ thereafter undergoes three-body
RPV decays driven by $\lambda''_{312}$, to give rise to final states
consisting leptons and jets of various multiplicities.

We observe that for the same values of ($m_0,m_{1/2}$), the mass and
branching fractions of the stop may vary drastically with different
values of ($\tan \beta, A_0$).  We shall choose $\mu > 0$ for all the
benchmark points as it is favoured by the constraint from the muon
anomalous magnetic moment\cite{Chattopadhyay:2001vx}. 

Since we explicitly want to study the situation in which the
neutralino decays within the detector, the only available decay mode
is $\tilde \chi_1^0 \rightarrow tds (\bar t \bar d \bar s)$. We
therefore require that the neutralino mass be greater than the top
mass to allow for a three-body decay.  We fix $m_{1/2}=450$ GeV which
gives $M_{\tilde \chi_1^0} \sim 180$ GeV.  We also choose the high
scale value of $\lambda''_{312}\sim 0.065$ such that it gives a value of
$0.2$ at the electroweak scale.

Figure \ref{bf} shows the branching fractions into various final
states for three different choices of $(\tan \beta, A_0)$, namely,
$(5,-1500)$, $(40,-1500)$ and $(10,0)$ for different stop masses,
obtained by varying $m_0$.  We notice that, for low $m_0$, the
dominant decay mode is $b\chi_2^+$ in the third case of Figure
\ref{bf}, while it is $t\chi_1^0$ in the first two cases.  We also
notice that the decays into higher neutralinos and charginos open up
earlier for $\tan \beta=$ 40 and compared to $\tan \beta=$ 5.

The Tevatron reach for single stop production is about $450$ GeV.  We
therefore start with a benchmark point with stop mass of 500 GeV, just
beyond this reach (Point A). The major decay channels in this case are
$t\tilde \chi_1^0,b\tilde \chi_1^+$.  A stop mass of a TeV at the
electroweak scale may be obtained by various configurations in the
high-scale parameter space.  However, from the above plots, one
expects its decays to change significantly with different parameters.
Our objective is to determine whether signal of resonant production of
a stop of mass near a TeV can be probed irrespective of what the
high-scale parameters are.  For this, we fix $M_{\tilde t_1} \sim 1$
TeV.  We first look at the case with $A_0=-1500$.  We construct two
benchmark points with $\tan \beta =$ 5(Point B) and 40(Point C) which
correspond to the opposite ends of the allowed range in $\tan \beta$.
We see that for a stop mass of 1 TeV, the decays into the
higgsino-like $\tilde \chi_2^+$ and $\tilde \chi_3^0$ become dominant
goes to high $\tan \beta$.

Similarly, we also look at a point with $A_0=0$ $\tan \beta =10$
(Point D).  In this case, we find that the Higgsino channels open up
fairly early and the dominant decay is $b\tilde \chi_2^+$ followed by
$t \tilde \chi_3^0$.  As we shall see in the next section, this plays
a crucial role in enhancing multi-lepton signals of a resonantly
produced stop.  Finally, since the decay into a top and a gluino does
not open up until much higher stop masses, we also construct one point
in which the stop decays dominantly into $t \tilde g$ (Point E).

Points A, B and E correspond to the same value of $(\tan \beta,
A_0)=(5,-1500)$ and therefore provide a description of how the signal
changes when only $m_0$ is varied.  This choice of parameters also
corresponds to the most conservative case in terms of signal since the
decay modes into the higher gauginos does not open for a large region
in the parameter space.  We will therefore use these points to obtain
limits on $\lambda''_{eff}$.

We have tabulated the parameters and significant decay modes in
Table~\ref{benchmarks}.  The benchmark points were generated with RPV
renormalisation group running of couplings and masses using SOFTSUSY
3.0.2\cite{Allanach:2009bv} and the RPV decays were calculated with
the ISAWIG interface to Isajet\cite{Paige:2003mg}.

\TABLE {
\begin{tabular}{|l|l|r|r|l|}
\hline
Point & $(m_0,\tan \beta,A_0)$ & $M_{\tilde t_1}$&$\sin^2\theta_{\tilde t}$ & Dominant decay modes \\
\hline
A &(600,5,-1500) &508 &0.88 &$t\tilde \chi_1^0$ (0.35); $b\tilde \chi_1^+$ (0.19); $\bar d \bar s$ (0.48) \\
\hline
B &(1650,5,-1500)& 1002 &0.97 &$t\tilde \chi_1^0$ (0.48);$t\tilde \chi_2^0$ (0.04); $b\tilde \chi_1^+$ (0.10); $\bar d \bar s$ (0.38) \\
\hline
C & (1570,40,-1500) & 1002 & 0.95 & $t \tilde \chi_1^0$ (0.35);$t \tilde \chi_2^0$ (0.05); \\
  & & & & $b \tilde \chi_1^+$ (0.12); $b \tilde \chi_2^+$ (0.21); $\bar d \bar s$ (0.27)\\
\hline
D & (1250,10,0) & 1008 &0.97  & $t \tilde \chi_1^0$ (0.13); $t \tilde \chi_2^0$ (0.04); $t \tilde \chi_3^0$ (0.20); $t \tilde \chi_4^0$ (0.13);\\
  & & & &$b \tilde \chi_1^+$ (0.08); $b \tilde \chi_2^+$ (0.33);  $\bar d \bar s$ (0.10) \\
\hline
E &(2450,5,-1500) & 1404 &0.99 & $t \tilde g$ (0.39); $t \tilde \chi_1^0$ (0.15); $t \tilde \chi_2^0$ (0.02); $t \tilde \chi_3^0$ (0.08); \\
  & & & &$t \tilde \chi_4^0$ (0.07); $b \tilde \chi_1^+$ (0.02); $b \tilde \chi_2^+$ (0.17);  $\bar d \bar s$ (0.11) \\
\hline
\end{tabular}
\label{benchmarks}
\caption{Benchmark points and the dominant decay modes of the lighter
  stop.  $\lambda''_{312}=0.2$, $\mu > 0$ and $m_{1/2}=450$~GeV for all
  benchmark points.}
}

The decay width of the stop in the R-parity violating channel $ds$
depends only on $\lambda''_{eff}$ and the stop mass.  Therefore, the
branching ratio into this channel for same values of $\lambda''_{eff}$
and stop mass depends only on the decay widths of the other channels
open at the same time.  For the benchmarks under consideration,
$\tilde \chi_{1,2}^0$ and $\tilde \chi_{1}^\pm$ have large gaugino
fractions whereas $\tilde \chi_{3,4}^0$ and $\tilde \chi_{2}^\pm$ have
large higgsino fractions.  The large top mass means that stop coupling
to higgsino-like chargino and neutralinos is large.  Thus as soon as
these decays become kinematically allowed, they quickly dominate over
the decays into gaugino-like chargino and neutralinos.  This can be
seen for points B, C and D which have nearly identical stop masses and
$\sin^2 \theta_{\tilde t}$ ($\lambda''_{312}=0.2$ at electroweak scale
for all points).  Large $\tan \beta$ opens up the $b \tilde \chi_2^+$
mode early in point C as compared to point B and makes the branching
fraction into $ds$ for point C much lower.  For point D, the branching
fraction into higgsino-like chargino and neutralinos is larger than
$60 \%$ and the RPV decay fraction is only about $10\%$.  The $\tilde
t-t-\tilde g$ coupling comes from strong interactions and therefore the
$t\tilde g$ channel dominates whenever it becomes kinematically
allowed (as in point E).

\section{Event generation and selection}

\subsection{Event generation}
Signal events have been generated using
HERWIG~6.510\cite{Corcella:2002jc}, and jets have been formed using
anti-$k_T$ algorithm\cite{Cacciari:2008gp} from FastJet~2.4.1.  SM
backgrounds have been calculated using
Alpgen~2.13\cite{Mangano:2002ea} showered through Pythia
\cite{Sjostrand:2006za} with MLM matching.  We have used CTEQ6L1
parton distribution functions\cite{Pumplin:2002vw}.  The
renormalisation and factorisation scales have been set at the lighter
stop mass ($M_{\tilde t_1}$) for signal, while the default option in
ALPGEN has been used for the backgrounds.

In R-parity conserving MSSM, the production of two heavy
superparticles requires a large centre-of-mass energy at the parton
level.  This allows us to further suppress the SM background by
applying cuts on global variables like the ``effective mass''
($M_{eff}$). Since we no longer have a large missing-$E_T$ and the
energy scale of the resonant production process is not very high, the
SM background cannot be suppressed so easily.  We therefore concentrate
on leptonic signals with or without b-tags to identify the signal over
the background.

\subsection{Event selection}
Decay of the lighter stop in this scenario can lead to a variety of
final states.  Out of them, we have chosen the following ones:
\begin{itemize}
\item Same-sign dileptons: $SSD$
\item Same-sign dileptons with one b-tagged jet: $SSD+b$
\item Trileptons: $3l$
\end{itemize}

We do not consider the RPV dijet channel as a viable signature due to
the enormous background from QCD processes.  Similarly, we also omit
opposite-sign dileptons due to large backgrounds from Drell-Yan,
$W^+W^-$, $t \bar t$ etc.

We have imposed the following identification requirements on leptons
and jets:

\begin{itemize}
\item \textbf{Leptons:} A lepton ($l$) is considered isolated if (a)
  It is well separated from each jet ($j$): $\Delta R_{lj}>0.4$, (b)
  The total hadronic deposit within $\Delta R<0.35$ is less than
  $10$~GeV.  We consider only those leptons which fall within
  $|\eta|<2.5$ with $p_T>10$~GeV.  Here, $\Delta R = \sqrt{\Delta
    \eta^2 + \Delta\phi^2}$ where $\eta$ is the pseudo-rapidity and
  $\phi$ is the azimuthal angle.
\item \textbf{Jets:} Jets have been formed using the anti-$k_T$
  algorithm with parameter $R=0.7$.  We only retain jets with
  $p_T>20$~GeV and $|\eta|<2.5$.
\item \textbf{B-tagged jets:} A jet is b-tagged with probability of
  $0.5$ if a b-hadron with $50<p_T<100$~GeV lies within a cone of
  $0.7$ from the jet axis.  We have set the identification efficiency
  to be zero outside this window, in order to make our estimates
  conservative.
\end{itemize}

\noindent We also apply the following extra cuts on various final
states to enhance the signal over background:
\begin{itemize}
\item \textbf{Cut 1: Lepton-$p_T$}: We demand that the $p_T$ of the
  leptons be greater than $(40,30)$ GeV for dilepton and $(30,30,20)$
  for trilepton channels.  This cut removes the background from
  semileptonic decays of b quarks.  It strongly suppresses the $b \bar
  b+jets$, $W b \bar b+jets$ and $t \bar t+jets$ background in the
  $SSD$-channel coming from semileptonic b-decays.
\item \textbf{Cut 2: Missing $E_T$}: At least one lepton in the signal always
  comes from the decay of a W boson and is accompanied by a neutrino.
  We demand a missing-$E_T$ greater than $30$~GeV from all events.
  This helps in reducing the probability of jets faking leptons.
  Missing-$E_T$ has been defined as $|\vec p_{T,visible}|$.
\item \textbf{Cut 3: Jet $p_T$}: We demand that the number of jets,
  $n_{j} \geq 2$ with $p_T(j_1)>100$ GeV and $p_t(j_2)>50.0$ GeV for
  $SSD$ and $SSD+b$.  This cut is useful when high stop mass is very
  high and the production cross section is very low.
\item \textbf{Cut 4: Dilepton invariant mass}: We also apply a cut on
  dilepton invariant mass ($M_{l1,l2}$) around Z-mass window
  ($|M_{l_1,l_2}-M_Z|< 15.0$ GeV) for opposite sign dileptons of same
  flavour in trilepton events.  This serves to suppress contribution
  from $Zb\bar b+jets$ and $WZ+jets$ background to trileptons.
\end{itemize}

Due to the Majorana nature of neutralinos, $\lambda''$-type
interactions result in equal rates for $tds$ and $\bar t \bar d \bar
s$-type final states.  Therefore, the most promising signals are those
involving same-sign dileptons (SSD).  This not only applies to $\tilde
\chi_1^0$ but also to the higher neutralinos produced in stop decay,
whose cascades can give rise to W's.  SSD have previously been used
extensively for studying signals of supersymmetry
\cite{Barnett:1993ea, Dreiner:1993ba} .  The most copious backgrounds
to SSD processes come from the processes $t \bar t$ and $Wb \bar b$
due to one lepton from $W$ and another from semileptonic decays of the
b-quark. There is also a potentially large contribution from $b \bar
b$ due to $B^0- \bar B^0$ oscillations along with semileptonic decays
of both B-mesons.  The effect of oscillations is simulated in the
Pythia program.  The $p_T$-cuts on leptons have been selected to
minimise the background from heavy flavour decays
\cite{Sullivan:2010jk}.  We find that after the isolation and $p_T$
cuts on leptons, $Wb \bar b$ and $b\bar b$ cross sections fall to
sub-femtobarn levels.

We simulate the $t \bar t + jets$ background up to two jets.  The
trilepton channel has another source of backgrounds in $WZ + jets$;
however, we have checked and found them to be negligibly small after
applying all the cuts.  We also generate $Wt\bar t+jets$ and $Zt\bar
t+jets$ up to one jet.

It should be mentioned here that the dilepton and trilepton final
states can also arise in the same scenario from the pair-production of
superparticles. These include, for example, pair production of gluinos
and electroweak production of chargino-neutralino pairs.  Such
contributions have been explicitly shown in the plots in section 4.

We also expect that the $p_T$ distribution of the $\tilde t_1$ becomes
significantly harder if the NLO corrections are taken into
account\cite{Plehn:2000be}.  Our cuts on leptons have been designed to
cut off the background from semileptonic b-decays by requiring the
$p_T$ to be about half the mass of the $W$.  Therefore, if only the
lepton cuts are used, we do not expect a large change in the
efficiency of the cuts quoted in the next section.

\section{Results}

We present results using $\lambda''_{312} = 0.2$; the predictions for
other values of this coupling can be obtained through scaling
arguments.  If $\lambda''_{312}$ is scaled by a factor of $n$ then the
production cross section as well as the decay width of the RPV channel
scale by $n^2$.  All other decay widths remain unchanged.  If $f$ is
the branching fraction of $\tilde t \rightarrow \bar d \bar s$ before
scaling, then the signal rates in any other channel are scaled by a
factor of

\begin{equation}
\frac{R_{new}}{R_{old}}=\frac{n^2}{(n^2-1)f+1}
\end{equation}

\subsection{Limits at $\sqrt{s}=$14, 10 TeV}

The numerical results for various signals corresponding to the five
benchmark points for LHC running energy $\sqrt{s} = 14,10$~TeV are
presented in Tables \ref{cutflow-ssd}(for the $SSD$ channel),
\ref{cutflow-ssdb} (for $SSD+b$) and \ref{cutflow-trilep} (for $3l$).

\TABLE{
\begin{tabular}{|l|cccc|cccc|}
\hline
 $SSD$  &           & 14 TeV    &           &          &           &  10 TeV   &           &          \\
 Point  &    Cut 0  &    Cut 1  &    Cut 2  &   Cut 3  &    Cut 0  &    Cut 1  &    Cut 2  &   Cut 3  \\
\hline
 A      &  884.8  &  496.8  &{\bf 459.4} &  41.0  &  540.1  &  312.7  &{\bf 287.0} &  15.1  \\
 B      &   64.7  &   43.7  &{\bf 41.4}  &  19.3  &   30.6  &   21.0  &{\bf  19.8} &   9.6  \\
 C      &   83.0  &   51.5  &{\bf 49.2}  &  25.8  &   40.1  &   25.6  &{\bf  24.6}  &  12.5  \\
 D      &  145.4  &   71.9  &{\bf 68.9}  &  41.1  &   65.1  &   32.3  &{\bf  31.0}  &  19.0  \\
 E      &   29.8  &   16.5  &   15.9  &{\bf 13.6}&  10.7  &    5.8  &    5.6  & {\bf 4.6} \\
\hline
 $t \bar t +nj$  & 687.9  &  26.3  &  24.7  &  10.0  &  307.0  &  8.7   &  7.0   &  3.6   \\
 $Wt \bar t +nj$ &  17.0  &   9.2  &   8.7  &   5.2  &    7.6  &  3.9   &  3.7   &  2.0   \\
 $Zt \bar t +nj$ &  12.7  &   6.7  &   6.7  &   4.1  &    4.9  &  2.3   &  2.2   &  1.4   \\
\hline
 Total           & 717.6  &  42.2  & 40.1   &  19.3  &  319.5  & 14.9   & 12.9   &  7.0   \\
\hline
\end{tabular}
\label{cutflow-ssd}
\caption{Effect of cuts on signal and SM background cross sections (in
  $fb$) in the $SSD$ channel at $\sqrt{s}=14,10$ TeV. Cut 0 refers to
  all events passing the identification cuts.  Cuts 1-3 are described
  in the text.  The numbers corresponding to best significance
  ($s/\sqrt{b}$) of the signal ($s$) with respect to the background
  ($b$) are highlighted in bold.}  }

\TABLE{
\begin{tabular}{|l|cccc|cccc|}
\hline
 $SSD+b$ &         & 14 TeV    &           &          &           &  10 TeV   &           &          \\
 Point   &  Cut 0  &    Cut 1  &    Cut 2  &   Cut 3  &    Cut 0  &    Cut 1  &    Cut 2  &   Cut 3  \\
\hline
 A       &  243.2  &  134.7 & \textbf{121.2} & 14.1   &  158.7 &  67.5 & \textbf{61.8} &  6.6     \\
 B       &  13.8   &  9.6   & \textbf{9.2}     &  4.7   &  8.3   &   6.0 & \textbf{5.6}  &  2.8     \\
 C       &  25.3   & 15.2   & \textbf{14.6}    &  8.0   & 12.7   &   7.8 & \textbf{7.4}  &  4.1     \\
 D       &  47.9   & 23.9   & \textbf{23.0}    & 15.2   & 21.2   &   9.9 & \textbf{9.5}  &  6.3     \\
 E       &  11.3   &  6.2   &   6.0     & \textbf{ 5.3} &  4.1   &   2.2 &          2.1  &\textbf{1.8} \\

\hline
 $t \bar t +nj$  & 173.0  &   7.6  &  7.1   &  2.3   &  80.9   &  4.2   &  1.4   &  1.3   \\
 $Wt \bar t +nj$ &  6.7   &   0.8  &   0.8  &   0.6  &  4.0    &  2.1   &  1.9   &  1.3   \\
 $Zt \bar t +nj$ &  5.6   &   2.6  &   2.6  &   1.8  &  2.3    &  1.1   &  1.1   &  0.7   \\
\hline
 Total           & 185.3  &  11.0  & 10.5   &   4.7  &  87.2   &  7.4   &  4.4   &  3.3   \\
\hline
\end{tabular}
\label{cutflow-ssdb}
\caption{Same as Table \ref{cutflow-ssd}, but for the $SSD+b$
  channel.}  
}

\TABLE{
\begin{tabular}{|l|cccc|cccc|}
\hline
 $3l$ &        &        & 14 TeV &        &       &       &10 TeV &         \\
 Point   & Cut 0   & Cut 1     & Cut 2   & Cut 4    & Cut 0 & Cut 1 & Cut 2   &  Cut 4 \\
\hline
 A       &   49.1  &   2.8     &  2.8    & 0.0      &  18.0     &  1.1      &  1.1      &  0.0     \\
 B       &    2.0  &  0.6      &  0.6    & 0.1      &  1.1      &  0.2      &  0.2      &  0.0     \\
 C       &   13.7  &  9.1      &  9.1    &  1.1     &  6.5      &  3.7      &  3.7      &  0.6     \\
 D       &   48.2  & 29.6      &  29.0   & 8.6      &  24.1     & 14.5      & 14.1      & 4.5      \\
 E       &   9.3   &  5.7      &   5.5   &  3.0     &  3.4      &  2.1      &  2.1      & 1.2      \\

\hline
 $t \bar t +nj$  &   2.1  &   0.0  &  0.0   &  0.0   &  1.8    &  0.0   &  0.0    &  0.0   \\
 $Wt \bar t +nj$ &  4.1   &   2.5  &  2.4   &  1.0   &  2.2    &  1.4   &  1.4   &  0.6   \\
 $Zt \bar t +nj$ & 30.8   &   20.7 & 19.7   &  2.7   & 11.3    &  7.3   &  4.7   &  1.1   \\
\hline
 Total           &  37.0  &  23.2  & 22.1   &   3.7  & 15.3    &  8.7   &  6.1   & 1.7     \\
\hline
\end{tabular}
\label{cutflow-trilep}
\caption{Effect of cuts on signal and SM background cross sections (in
  $fb$) in the trilepton ($3l$) channel at $\sqrt{s}=14,10$ TeV. Cut 0
  refers to all events passing the identification cuts.  Cuts 1, 2 and
  4 are described in the text. Cut 4 is necessary to eliminate
  background from $WZ+jets$.}  }

We can make the following observations from the numerical results:
\begin{itemize}
\item The final states $SSD$ and $SSD + b$ consistently have
  substantial event rates at both 14 and 10 TeV.  Furthermore, the
  simultaneous observation of excesses in the SSD and SSD +b channel
  can serve as definite pointer to the production of a third
  generation squark.

\item For point A, which is just above the Tevatron reach, we can
  achieve more than $5\sigma$ significance in the $SSD$ channel with
  just 100 pb$^{-1}$ data at both 14 and 10 TeV.  For point E, which
  has $M_{\tilde t}=1500$ GeV, we can reach $3\sigma$ with 1(3)
  $fb^{-1}$ and $5\sigma$ with 3(9) $fb^{-1}$ at 14(10) TeV.
  Therefore, we can conclude that the entire range from 500-1500 GeV
  can be successfully probed at the LHC for $\lambda''_{312}=0.2$.

\item Stops decaying into higher neutralinos and charginos make the
  total rates distinctly better.  This is governed by Higgsino
  couplings and is therefore most prominent for high $\tan \beta$ and
  low $A_0$.  This effect is evident from the large event rates for
  point D.  We can successfully probe this point in the SSD channel at
  $5\sigma$ with less than 1 $fb^{-1}$ data at both 10 and 14 TeV
  runs.

\item The trilepton final state occurs when the stop can decay into
  $\chi_2^+$, $\chi_{3,4}^0$ or $\tilde g$.  Therefore, points A and B
  show almost no signal and Point D has the largest signal in this
  channel.  This advantage is largely lost for benchmark point E due
  to the kinematic suppression in the stop production process.

\item Reach for the LHC: Assuming the conservative case of ($\tan
  \beta =5,~A_0=0$), with 10 $fb^{-1}$ luminosity, one can rule out
  $\lambda_{eff}^{''}$ greater than 0.007--0.045 (0.007--0.062) for
  stop masses between 500 and 1500 GeV at 95 \% CL at $\sqrt{s}=$
  14 (10) TeV.  A $5\sigma$ discovery can be made in the same mass
  range for $\lambda''_{eff}$ greater than 0.012--0.084 (0.012--0.12).
  However, we observe that the reach in stop mass does not decrease
  monotonously with stop mass.  The opening of new decay channels can
  improve detection considerably.  The statements about minimum value
  of $\lambda_{eff}^{''}$ that can be probed are therefore dependent
  on the particular decays of the stop.  We therefore tabulate the
  minimum values of $\lambda_{eff}^{''}$ for each benchmark point at
  10 $fb^{-1}$ for both 10 and 14 TeV in Table \ref{min-lam}.
\end{itemize}

\TABLE{
\begin{tabular}{|l|rrr|rrr|}
\hline
 Point  &           &  14 TeV  &         &            &  10 TeV  &         \\
\hline
        &  95\% CL  & $3\sigma$&$5\sigma$&  95 \% CL  &$3\sigma$ &$5\sigma$ \\
\hline
 A      &    0.007  &   0.009  &  0.012  &     0.007  &   0.009  &  0.012  \\
 B      &    0.027  &   0.037  &  0.052  &     0.029  &   0.041  &  0.059  \\
 C      &    0.026  &   0.035  &  0.048  &     0.028  &   0.038  &  0.052  \\
 D      &    0.024  &   0.032  &  0.042  &     0.027  &   0.036  &  0.047  \\
 E      &    0.045  &   0.062  &  0.084  &     0.062  &   0.087  &   0.12  \\
\hline
\end{tabular}
\caption{Values of minimum $\lambda_{eff}''$ that can be ruled out at
  95\% CL, probed at $3\sigma$ or $5\sigma$ with 10 $fb^{-1}$ of data
  at $\sqrt{s}=10,14$ for each of the benchmark points.  The
  significance used is $s/\sqrt{b}$ where $s$ is the signal and $b$ is
  the background. $s/b>0.2$ in all cases.}
\label{min-lam}
}

In Figure \ref{ssd-E}, we present the effective mass distributions in
$SSD$ channel for all the benchmark points. Effective mass is defined
as
\begin{equation}
M_{eff}=\sum_{jets}|\vec p_T|+\sum_{leptons}|\vec p_T|+E_T\miss
\label{meff}
\end{equation}

The contributions from resonant stop production is superposed in the
figures on the SM backgrounds and also RPC superparticle production
processes.  The RPC contributions are much smaller and therefore do
not provide a serious background to our signals.

\FIGURE{
\includegraphics[width=70mm]{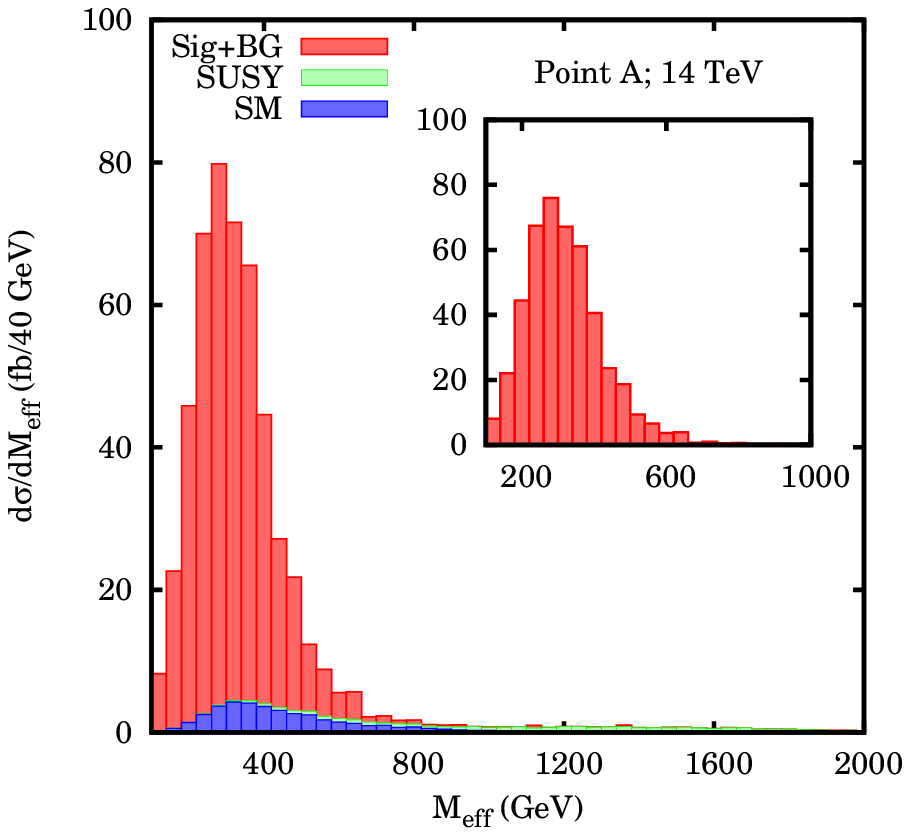}

\includegraphics[width=70mm]{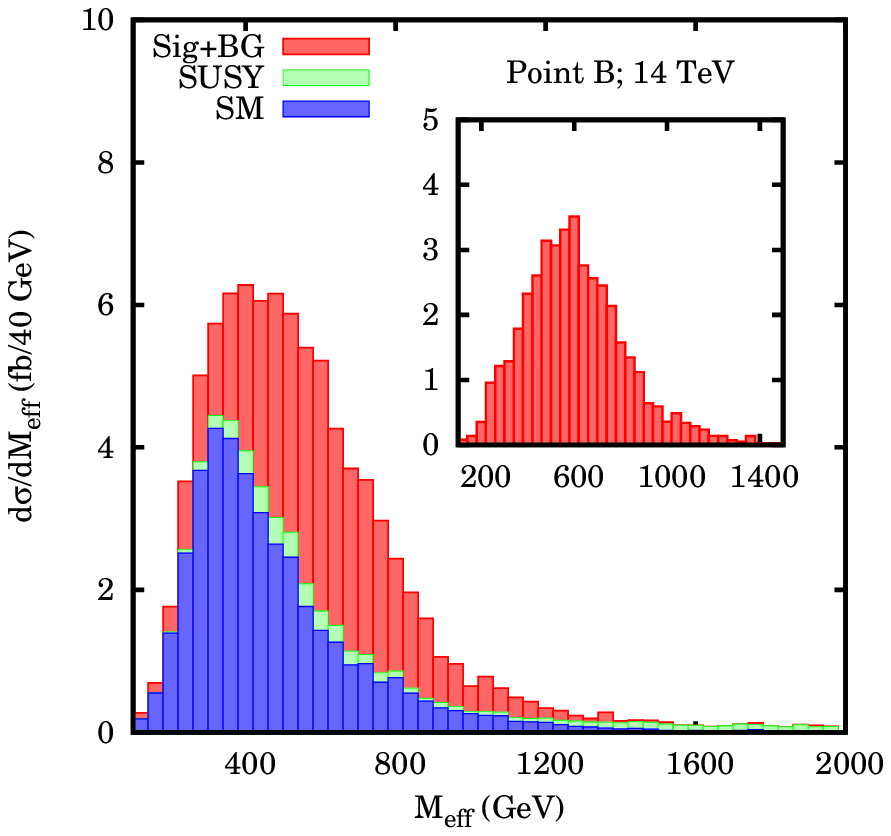}

\includegraphics[width=70mm]{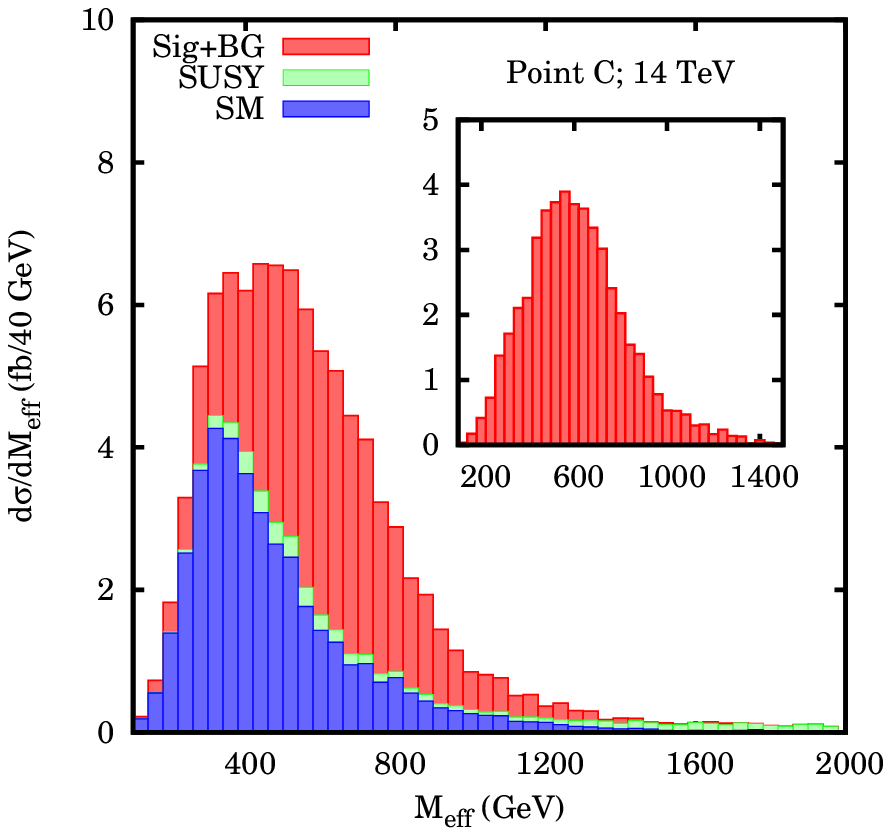}

\includegraphics[width=70mm]{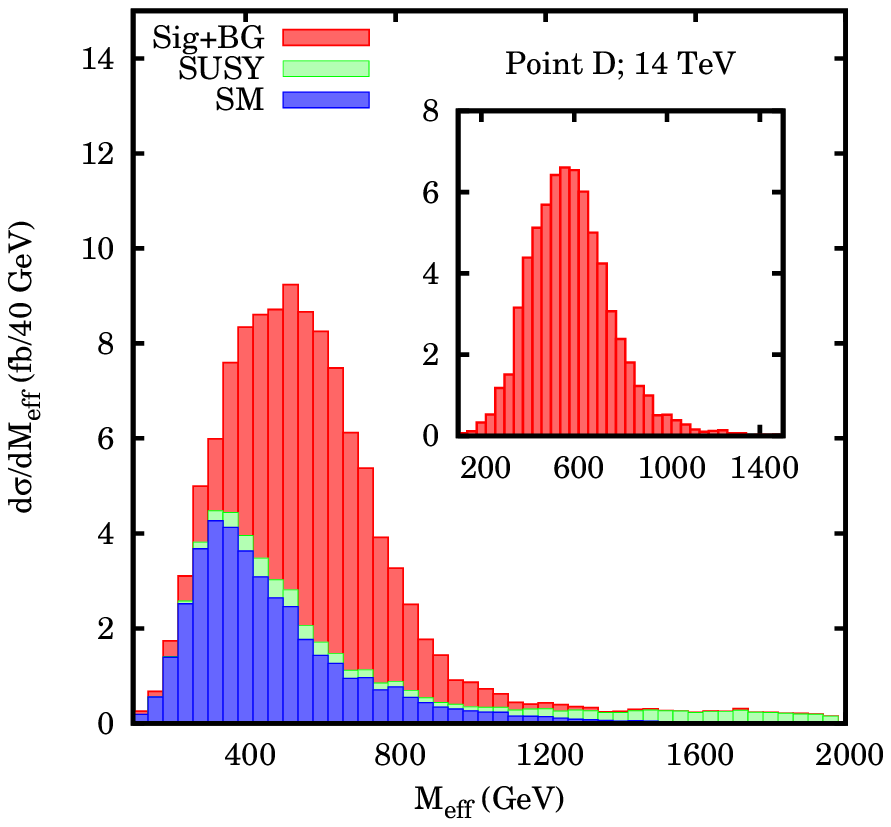}

\includegraphics[width=70mm]{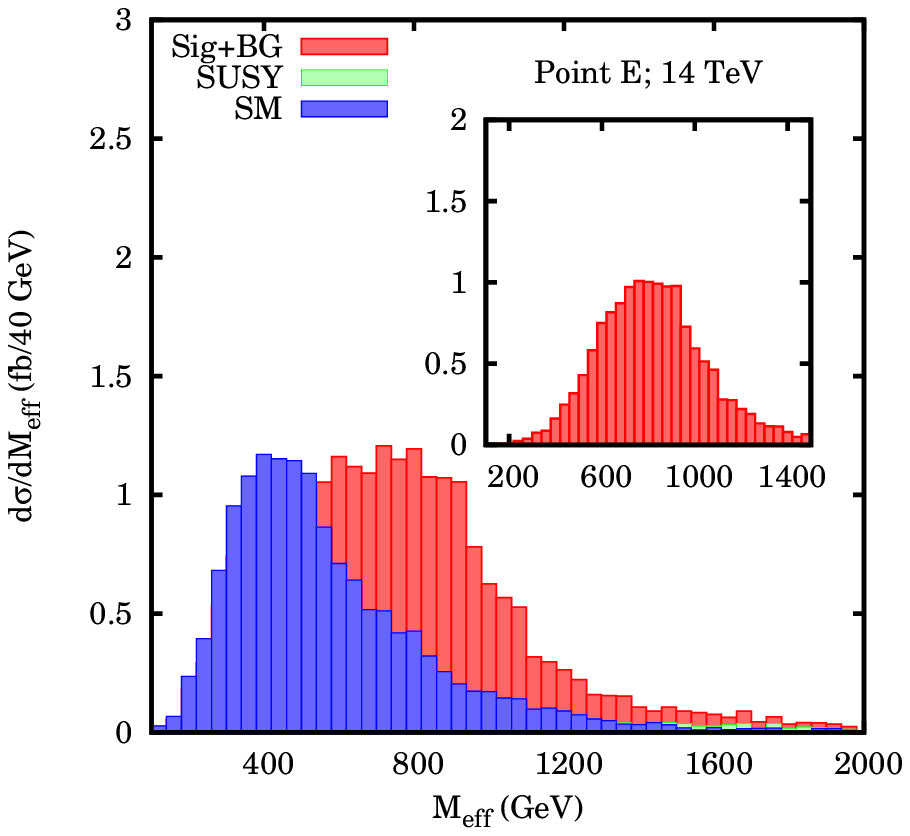}

\caption{$M_{eff}$ distributions at $\sqrt s = 14$~TeV.  ``SM'' is
  the contribution to the background from Standard Model processes.
  ``SUSY'' refers to the contribution from R-conserving production
  processes.  The inset in each figure contains the distribution for
  the signal alone.}
\label{ssd-E}
}

\subsection{Observability at the early run of 7 TeV}

The initial LHC run at $\sqrt{s} =$ 7 TeV will collect up to
$1~\mathrm{fb}^{-1}$ data.  It will be difficult to observe RPV
production of a 1 TeV stop at this energy.  However, we can make
useful comments for lower stop masses by looking at the $SSD$ channel.
We therefore look two benchmark points with low stop masses: the first
is the `Point A' described earlier and the second is similar to `Point
D' (with $\tan \beta=10$ and $A_0=0$).  Since the $t\bar t$
backgrounds are much smaller at 7 TeV, we relax the jet-$p_T$ cuts.
The high scale parameters, stop mass at electroweak scale and cut-flow
table for signal as well as background are given in Table
\ref{cutflow-7tev}.

We conclude that we can rule out up to $\lambda''_{eff}=0.025$ for a
stop mass of 500 GeV at 7 TeV with $1~\mathrm{fb}^{-1}$ data and a
$5\sigma$ discovery can be made at stop mass 500 GeV for
$\lambda''_{eff} \geq 0.043$.  For the case $\tan \beta=10$ and
$A_0=0$, the lowest possible theoretically allowed stop mass (with
$m_{1/2}=450$~GeV) is 775 GeV and we can rule out up to
$\lambda''_{eff}=0.054$ with $1~\mathrm{fb}^{-1}$ data.

\TABLE{
\begin{tabular}{|l|cc|ccc|}
\hline
 Point & $(\tan \beta, A_0, m_0)$ & $M_{\tilde t_1}$ & Cut 0  & Cut 1  & Cut 2   \\
\hline
$A$   & $(5,-1500,600)$   & 508 & 283.3 & 158.6  & 147.5 \\
$D^\prime$ & $(10,0,100)$      & 775 &  70.0 &  33.9  & 32.0  \\
\hline
 $t \bar t +nj$  &        &     & 116.0 &  3.7   & 3.5   \\
 $Wt \bar t +nj$ &        &     & 4.3   &  2.3   & 2.1   \\
 $Zt \bar t +nj$ &        &     & 1.8   &  0.9   & 0.8   \\ 
\hline
Total            &        &     & 122.1 &  6.9   & 6.4   \\
\hline
\end{tabular}
\label{cutflow-7tev}
\caption{The benchmark points for studying RPV stop production and the
  effect of cuts on signal and SM background cross sections (in $fb$)
  in the $SSD$ channel at $\sqrt{s}=7$ TeV. Cut 0 refers to all events
  passing the identification cuts.  All other cuts are described in
  the text, we do not apply Cut 3.}  }

\subsection{Differentiating from R-conserving signals}
We now address the question whether the signals we suggest can be
faked by an R-parity conserving scenario in some other region(s) of
the parameter space.  One possible way that our signal may be mimicked
is if a point in the mSUGRA parameter space (without RPV) gives
similar kinematic distributions to any our benchmark points.  More
specifically, one may have a peak in the same region for the variable
$M_{eff}$, defined in equation \ref{meff}.  

For each of our benchmark points A-D, an $M_{eff}$ peak in the same
region requires the strongly interacting sparticles to have masses in
the range already ruled out by the Tevatron data\cite{Amsler:2008zzb}.
In particular, they require the gluino mass $M_{\tilde g}<390$
GeV. Thus the question of faking arises only for benchmark point E,
which represents the highest mass where the signals rates are
appreciable.

We generate such a point (Point RC) with the parameters $m_0=300$,
$m_{1/2}=180$, $A_0=0$, $\tan \beta= 10$, $\mu>0$ and the resultant
sparticle masses for coloured particles are $M_{\tilde g} =465$,
$M_{\tilde q} \sim 500$ GeV.  The $M_{eff}$ distributions for point E
and point RC is shown in Figure \ref{fake}.  We present the following
results at 14 TeV as an illustration.  Distributions at 10 TeV are
almost identical.

The missing $E_T$ distribution is also not a good discriminator under
such circumstances, as can be seen from Figure \ref{fake}.  This is
because the neutrinos that contribute to missing-$E_T$ in the RPV case
are highly boosted due to the large masses of the particles produced
in the initial hard scattering.  Thus, the $E_T$ spectrum is actually
harder for the RPV case even though the RPC case has a stable massive
LSP.  However, as the resultant spectrum is quite light, the RPC
production cross section ($\sim$ 40 pb) is about two orders of
magnitude greater than the RPV case with $\lambda''_{312}=0.2$
($\sim$480 fb).  Consequently, the rate of the SSD signals, for
example, are much higher in the R-conserving scenario ($\sim$ 34 fb)
as compared to those from point E ($\sim$ 14 fb).  For values of
$\lambda''_{312}<0.2$, we can therefore make a reliable distinction
simply based on the number of events expected in the SSD channel.

Another possible discriminator is the charge asymmetry.  In the $SSD$
channel, one can look at the ratio of negative to positive $SSD$
$\frac{N^{--}}{N^{++}}$.  The fraction of $ds \rightarrow \tilde
t_1^*$ is more than the charge conjugate process $\bar d \bar s
\rightarrow \tilde t_1$ due to the difference in parton distributions
of $d$ and $\bar d$ in the proton.  Therefore, one expects extra
negative sign leptons than positive ones. Whereas in the RPC case,
since most of the $SSD$ contribution comes from $\tilde g \tilde g$
production, we do not expect a large asymmetry.  In our illustration,
we see that this ratio is 2.7 (1.4) for the RPV (RPC) case.

\FIGURE{
\includegraphics[width=70mm]{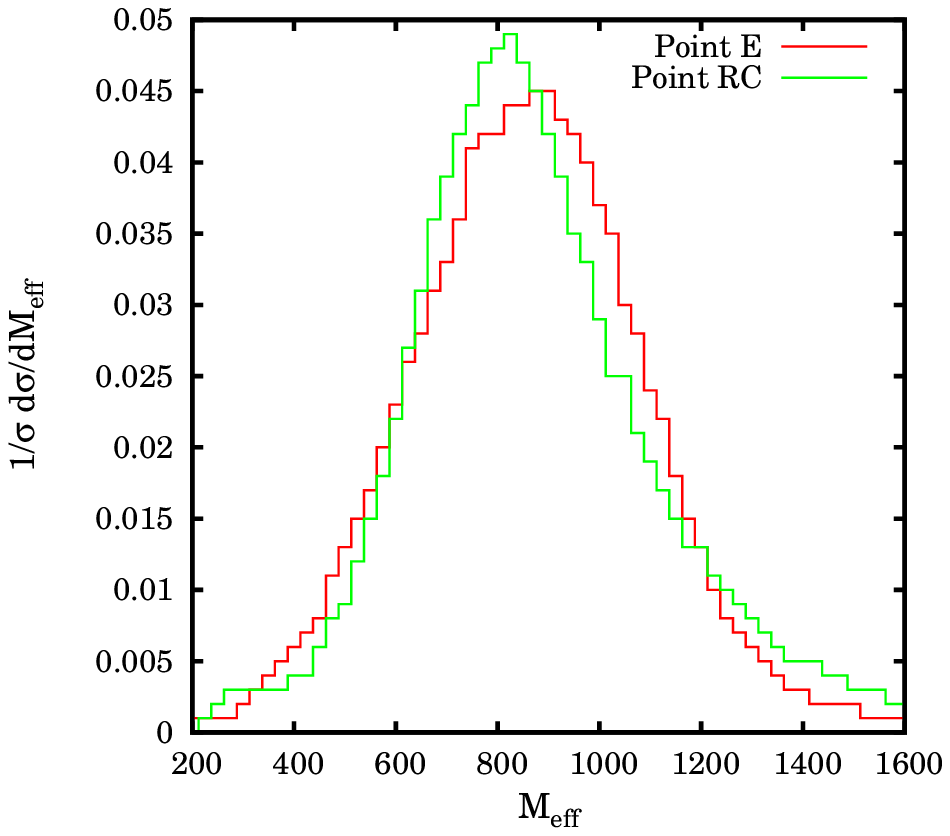}
\includegraphics[width=70mm]{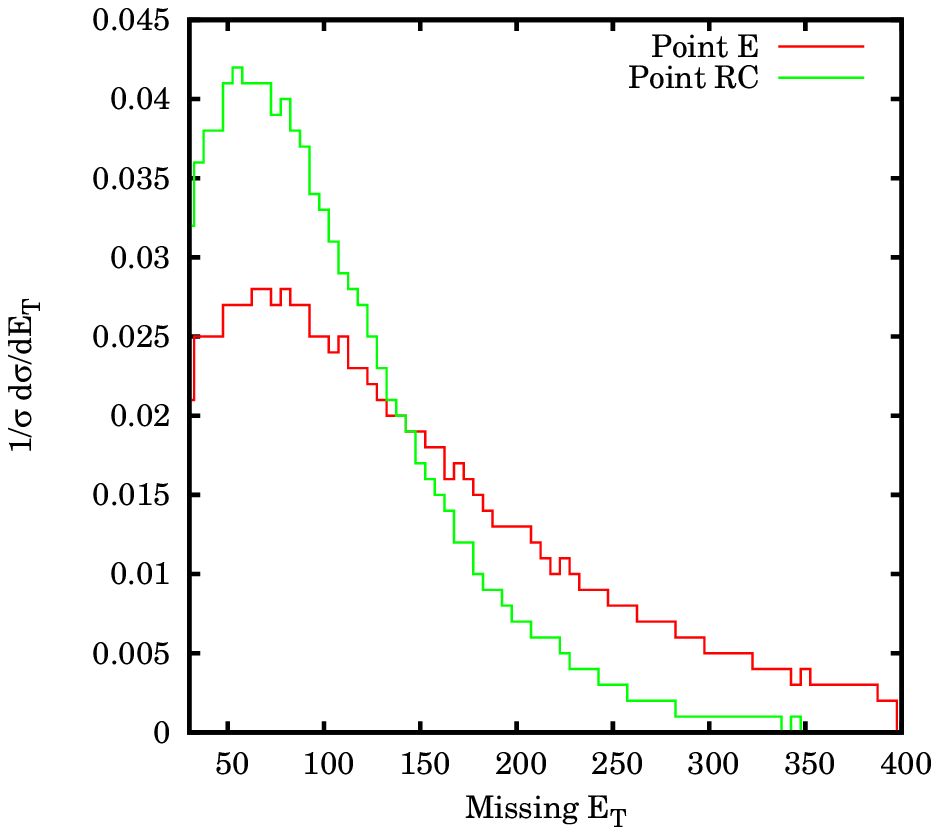}
\label{fake}
\caption{The normalised effective mass ($M_{eff}$) and missing energy
  ($E_T\miss$) distributions in the $SSD$ channel for Point E and an
  Point RC.  Point RC has been generated using $m_0=300,
  m_{1/2}=180;A_0=0, tan \beta=10$ and $\mu>0$.  The gluino mass is
  465 GeV.}  }

\section{Non-$\tilde \chi_1^0$ LSPs}
For RPV models, the restriction of having an uncharged LSP no longer
exits.  A significant region of the mSUGRA parameter space with low
$m_0$ corresponds to a stau ($\tilde \tau$) LSP.  With only
$\lambda''_{312}$-type couplings present, the stau can only decay via
off-shell $\tilde \chi_1^0$ and $\tilde t$ propagators into the four
body decay ($\tilde \tau \rightarrow \tau t d s$) if its mass,
$m_{\tilde \tau}>m_{top}$ or via the five body decay ($\tilde \tau
\rightarrow \tau b W d s$) if $m_{\tilde \tau}<m_{top}$ where the top
propagator is also off-shell.  The four-body decays of the stau in
lepton-number violating scenarios was calculated in
\cite{Dreiner:2008rv}.  Since the intermediate $\tilde \chi_1^0$ is of
Majorana character, we can always have one lepton of either sign from
LSP decay via the $W$ from an on-shell or off-shell top.  Thus, for
various types of $\tilde t$ decays, the following situations may
arise:
\begin{itemize}
\item For decays of $t\tilde \chi_i^0$-type, we can still have same-sign
  dileptons with one lepton from top decay and the other from the
  decay of the LSP. 
\item For decays of type $b \tilde \chi_i^\pm$ with $\chi_i^\pm
  \rightarrow W^\pm \tilde \chi_1^0 + X$, we have $\tilde \chi_1^0
  \rightarrow \tau \tilde \tau$ and the SSD come from $W^\pm$ and LSP
  decay respectively.
\item For decays of the type $\tilde t \rightarrow b \tilde \chi_i^+$
  with $\tilde \chi_i^+ \rightarrow \nu_{\tau} \tilde \tau + X$, we
  may still get SSD from leptonic decay of the $\tau$ in the $\tilde
  \tau$ decay.  If $\tau$-identification is used, final states of the
  type same-sign $(\tau + e /\mu)$ may be considered.
\item Since the stau has to decay via four- or five-body processes, it
  is possible that the lifetime of the $\tilde \tau$ is large and it is
  stable over the length scale of the detector.  In this case, it will
  leave a charged track like a muon and one can look at same-sign
  leptons with this ``muon'' as one of the leptons.  It is also
  possible that the lifetime is large but the stau still decays within
  the detector.  In this case, a displaced vertex can be observed in
  the detector.
\end{itemize}
We leave the detailed simulation of all scenarios of resonant stop
production with stau LSP to a future study.

Another possibility that arises with a large $\lambda''_{312}$ is of
having a stop LSP.  In this case however, the decay will be almost
entirely via the RPV $\bar d \bar s$ di-jet channel.  The
overwhelmingly large dijet backgrounds at the LHC would most likely
make this situation unobservable.

\section{Summary and Conclusion}

We have performed a detailed analysis of resonant stop production at
the LHC, both for the 10 and 14 TeV runs, for values of the baryon
number violating coupling $\lambda''_{312}$ an order of magnitude
below the current experimental limit.  Benchmark points have been
chosen for this purpose, which start just beyond the reach of the
Tevatron and end close at the LHC search limit.  We find that the
same-sign dilepton final states, both with and without a tagged b, are
most helpful in identifying the signal. The trilepton signals can also
be sometimes useful, especially when decays of the resonant into
higher neutralinos, the heavier chargino or the gluino open up. At
14(10) TeV, we can probe stop masses up to 1500 GeV and values of
$\lambda''_{eff}$ down to 0.05(0.06) depending on the combination of
various SUSY parameters. For cases of stop mass below a TeV, the
effective mass distributions can enable us to distinguish between the
resonant process and contributions from R-parity conserving SUSY
processes.  For higher stop masses, one has to rely on cross sections
or the charge asymmetry.

We have used a particular B-violating coupling, namely,
$\lambda''_{312}$.  One can also have resonant stop production driven
by $\lambda''_{313}$ and $\lambda''_{323}$.  In either of these cases,
one expects a larger abundance of b quarks in the final state.
However, there is a suppression in the production rates due to the
b-distribution function in the proton.

In conclusion, resonant stop production is a potentially interesting
channel to look for SUSY in its baryon-number violating
incarnation. Values of the B-violating coupling(s) more than an order
below the current experimental limits can be definitely probed at the
LHC, both at 10 and 14 TeV.  If such interactions really exist, our
suggested strategy can not only yield detectable event rates but also
point towards resonant production as opposed to pair-production of
SUSY particles.

\section*{Acknowledgements}
We would like to thank Satyaki Bhattacharya, Gobinda Majumdar, Bruce
Mellado and Satyanarayan Mukhopadhyay for helpful discussions and
comments.  ND thanks the Indian Association for the Cultivation of
Science for hospitality while part of this work was carried out.  This
work was partially supported by funding available from the Department
of Atomic Energy, Government of India for the Regional Centre for
Accelerator-based Particle Physics, Harish-Chandra Research
Institute. Computational work for this study was partially carried out
at the cluster computing facility of Harish-Chandra Research Institute
(\url{http://cluster.mri.ernet.in}).

\bibliographystyle{JHEP}
\bibliography{rpv-stop}

\end{document}